\documentstyle[epsf,floats,aps,prd,eqsecnum]{revtex}
\begin{document}
\title{Stochastic description for open quantum systems}
\author{Esteban Calzetta}
\address{Departamento de F\'{\i}sica,\\
Universidad de Buenos Aires, Ciudad Universitaria,\\
1428 Buenos Aires, Argentina}
\author{Albert Roura and Enric Verdaguer \thanks{
Also at Institut de F\'\i sica d'Altes Energies (IFAE), Barcelona, Spain.}}
\address{Departament de F\'{\i}sica Fonamental,\\
Universitat de Barcelona, Av.~Diagonal 647,\\
08028 Barcelona, Spain}
\maketitle

\begin{abstract}
A linear open quantum system consisting of a harmonic oscillator linearly
coupled to an infinite set of independent harmonic oscillators is
considered; these oscillators have a general spectral density function and
are initially in a Gaussian state. Using the influence functional formalism
a formal Langevin equation can be introduced to describe the system's fully
quantum properties even beyond the semiclassical regime. It is shown that
the reduced Wigner function for the system is exactly the formal
distribution function resulting from averaging both over the initial
conditions and the stochastic source of the formal Langevin equation. The
master equation for the reduced density matrix is then obtained in the same
way a Fokker-Planck equation can always be derived from a Langevin equation
characterizing a stochastic process. We also show that a subclass of quantum
correlation functions for the system can be deduced within the stochastic
description provided by the Langevin equation. It is emphasized that when
the system is not Markovian more information can be extracted from the
Langevin equation than from the master equation.
\end{abstract}

%\date{\today}
%\draft

\section{INTRODUCTION}

Feynman observed long ago that the dynamics of an open quantum system may be
described in terms of an equivalent stochastic problem \cite{feynman63}. In
this paper, we elaborate on this insight by showing that a certain class of
quantum Green functions may be obtained directly as ensemble averages in the
stochastic formulation. Unlike earlier treatments of the origin of
stochasticity in the semiclassical limit of quantum theories \cite
{gell-mann93} we do not assume decoherence; rather, we are concerned with
the full quantum dynamics of the system.

The stochastic treatment follows from the observation that the Wigner
function of the open quantum system may be represented as the formal
distribution function resulting from averaging the solutions of an
appropriate Langevin equation both over the initial conditions and the
stochastic source. This representation, which is exact for linear systems,
may be extended to nonlinear cases through perturbative methods \cite{CRV01}.

We compare the main approaches to the analysis of the stochastic dynamics,
namely the Langevin and Fokker-Planck equations, to the corresponding
quantum approaches, namely the master equation and the Wigner function. We
show that the Fokker-Planck equation is the transport equation for the full
Wigner function, and as such it is equivalent to the master equation. The
Langevin equation, on the other hand, provides a more detailed description
of the dynamics, in the sense that the class of correlation functions which
may be retrieved from the Langevin equation is larger than the corresponding
class for the master or Fokker-Planck equations unless the dynamics is
Markovian.

The relationship between these different approaches is summarized in the
diagram of Fig. 1. The most fundamental description of the open quantum
system is that provided by the Feynman-Vernon influence functional \cite
{feynman63}. From this, we may derive the master equation, which gives the
dynamics of the reduced density matrix. The integral transform linking the
reduced density matrix to the reduced Wigner function allows us to convert
the master equation for the former into a transport or Fokker-Planck
equation for the latter. Feynman's insight that the influence functional may
be thought as well as an ensemble average over an equivalent stochastic
noise allows us to retrieve some correlation functions of the quantum
problem directly in terms of stochastic averages, while the Fokker-Planck
equation associated to the corresponding Langevin equation gives back the
master equation. Thus, the Langevin equation is a very useful tool to gain
information on the quantum properties of the system even beyond the
semiclassical regime, when it no longer describes the actual trajectories of
the system.

\begin{figure}[htb]
\centering
\leavevmode \epsfysize=6cm \epsfbox{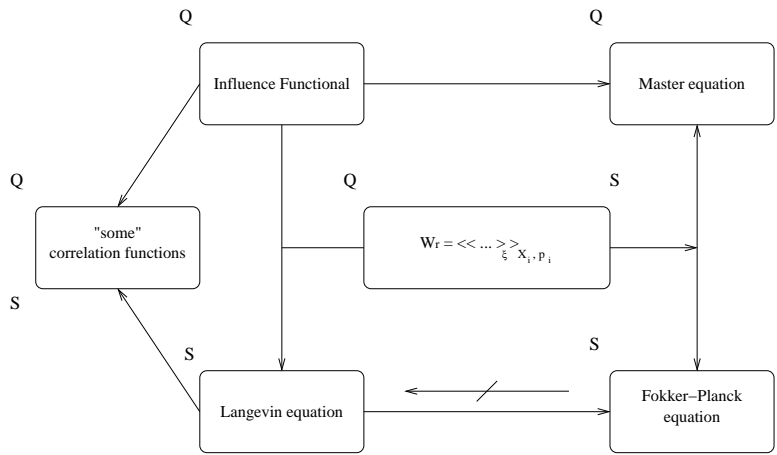}
\vspace{2cm}
\caption[fig1]{Diagram showing the interconnections between different
quantum properties of an open quantum system on the one hand, and between
the elments of the stochastic description on the other hand, as well as the
connection between both levels of description. Quantum objects (denoted by
label Q) are placed at the top of the diagram whereas those objects
associated to the stochastic description (denoted by label S) are placed at
the bottom. The equivalence of the reduced Wigner function with a
distribution function in the framework of the stochastic description
provided by the Langevin equation reflects the connection formally
established between both levels of description. One can take advantage of
this connection by working within the stochastic framework to obtain
relevant information about quantum properties. In the stochastic context,
one can always obtain a Fokker-Planck equation for the distribution function
associated to the stochastic process resulting from a given Langevin
equation. From the connection previously stated, the Fokker-Planck equation
can be translated into the master equation for the reduced Wigner function
or, equivalently, for the reduced density matrix. Similarly, correlation
functions for the stochastic processes which are solutions of the Langevin
equation are closely related to quantum correlation functions for system
observables.}
\label{fig1}
\end{figure}

The plan of the paper with a summary of the relevant results is the
following. In Sec.~\ref{sec2} we briefly review the quantum Brownian motion
(QBM) model describing a harmonic oscillator coupled bilinearly to a bath of
infinite harmonic oscillators initially in a Gaussian state and with an
arbitrary spectral distribution (general environment). We summarize the main
formulas that beginning with the influence functional lead to the reduced
density matrix operator and to the master equation. We also give the
equivalent evolution equation for the reduced Wigner function. These results
are well known and the computational details, which can be found in the
references provided, are omitted.

In Sec.~\ref{sec3} we derive one of the main results of this paper. We show
that the reduced Wigner function can be written as a formal distribution
function for the system variables defined as follows. We consider the
ensemble of systems in phase space which obey the formal Langevin equation
for a given realization of the stochastic force and arbitrary initial
conditions; the average both over the initial conditions (which involves the
initial reduced Wigner function) and the different realizations of the
stochastic source gives just the reduced Wigner function. The key technical
point that makes this result relatively easy to derive is the computation of
the path integral defining the reduced Wigner function in terms of variables
which include the initial conditions and the stochastic force; for this we
need to introduce a functional change which involves the formal Langevin
equation. The details of this derivation can be found in Appendix \ref{appA}%
. As far as we know this result is new. We are aware of a related result by
Halliwell and Zoupas \cite{halliwell95} in the limit of large times; see
also Ref. \cite{anglin96}.

Having shown that the reduced Wigner function is a formal distribution
function, it is clear that the dynamical equation for the reduced Wigner
function can be deduced using the usual techniques employed to derive the
Fokker-Planck equation \cite{sancho80}. We recall that whereas a Langevin
equation describes how an individual system evolves stochastically, a
Fokker-Planck equation describes how the distribution for an ensemble of
systems evolves deterministically in phase space. The details of this
derivation are left to Appendix \ref{appB}. This constitutes an alternative
derivation of the master equation to those given by Hu, Paz and Zhang \cite
{hu92} (see also Ref. \cite{paz94}), and Halliwell and Yu \cite{halliwell96}%
. We also note that this process is not reversible in the sense that there
may be many Langevin equations that give the same master equation. This just
reflects the fact that when the dynamics is not Markovian (in the sense of
Ref. \cite{paz93a}), more information can be extracted from the Langevin
equation than from the master equation.

In Sec.~\ref{sec4} we obtain the second main result of this paper. We show
that quantum correlation functions for the system variables can be obtained
within the stochastic description provided by the Langevin equation by
explicitly computing the closed time path (CTP) generating functional for
the system. It turns out that this generating functional can be written as
an average over the initial conditions times a term that depends on the
noise kernel, which contains the information on the fluctuations induced by
the environment on the system.

We also show that quantum correlation functions cannot be obtained using the
propagators for the reduced density matrix unless the system is Markovian, a
fact which is discussed in Appendix \ref{appC}. Note that this is in
contrast with a closed system where the unitary propagators, which are
solutions of the Schr\"{o}dinger equation, allow one to obtain all the
information about the existing quantum correlations.

Finally, in Sec.~\ref{sec5} we summarize and briefly discuss our results.

Throughout the paper we use units in which $\hbar =1$ except for Sec.~\ref
{sec2}.

\section{INFLUENCE FUNCTIONAL FORMALISM FOR OPEN QUANTUM SYSTEMS}

\label{sec2}

\subsection{A survey of open quantum systems}

Open quantum systems \cite{davies76} are of interest in condensed matter
physics \cite{caldeira83b}, quantum optics \cite{walls94}, quantum
measurement theory \cite{zurek81}, nonequilibrium field theory \cite
{calzetta88,calzetta00a,stephens99,calzetta00b}, quantum cosmology \cite
{habib90apaz91} and semiclassical gravity \cite{hu89}. Among the most widely
used examples of open quantum system is the QBM model, which consists of a
single massive particle in a potential (usually quadratic) interacting with
an infinite set of independent harmonic oscillators which are initially in a
Gaussian state (most often a thermal equilibrium state) \cite{zwanzig59}.
The coupling may be linear both in the system and environment variables or
may be nonlinear in some or all of these variables. The frequencies of the
environment oscillators are distributed according to a prescribed spectral
density function, the simplest case corresponding to the so-called ohmic
environment. The linear coupling provides a good description of many open
quantum systems in condensed matter physics \cite{caldeira83a,caldeira83b},
but in field theory \cite{calzetta97b}, quantum cosmology \cite
{habib90apaz91} and semiclassical gravity \cite{calzetta94,calzetta99} the
coupling is usually nonlinear. Part of the interest of the linear systems is
that they are in many cases exactly solvable and detailed studies of
different aspects of open quantum systems can be performed. One of the
issues that has received much attention in recent years is
environment-induced decoherence as a mechanism to understand the transition
from the quantum to the classical regime \cite{zurek91zurek93a,paz01}.

Concepts such as the Feynman-Vernon influence functional method, the reduced
density matrix, the reduced Wigner function, the master equation, the
Fokker-Planck equation and the Langevin equation are some of the key words
associated to the study of open quantum systems. One of the purposes of this
paper is to review the place of these concepts in the QBM model and to
establish their often subtle interrelations. Thus, let us first review some
of those concepts and recall their main features.

The {\em reduced density matrix} is defined from the density matrix of the
whole closed system by tracing out the environment. Its dynamical evolution
may be given in terms of the Feynman and Vernon influence functional \cite
{feynman63}. The influence functional is defined from a path integral
involving the action of the system and the environment and an integration of
the environment degrees of freedom. Its use in the QBM model is widespread
especially since Caldeira and Leggett were able to compute in closed form
the propagator for the reduced density matrix in the case of linear coupling
with an ohmic environment \cite{caldeira83a}.

The {\em master equation} is a differential equation describing the
evolution of the reduced density matrix. The master equation for linear
coupling and ohmic environment at high temperature was first deduced by
Caldeira and Leggett \cite{caldeira83a}, it was extended to arbitrary
temperature by Unruh and Zurek \cite{unruh89}, and it was finally obtained
for a general environment ({\em i.e.} for an arbitrary spectral density
function) by Hu, Paz and Zhang \cite{hu92}. This result can be extended to
the case of nonlinear coupling by treating the interaction perturbatively up
to quadratic order \cite{hu93}.

Closely related to the reduced density matrix is the {\em reduced Wigner
function} (in fact one goes from one to the other by an integral transform) 
\cite{wigner32,hillery86}. The reduced Wigner function is similar in many
aspects to a distribution function in phase-space, although it is not
necessarily positive definite, and the dynamical equation it satisfies is
similar to the {\em Fokker-Planck equation} for classical statistical
systems \cite{wax54,risken89}. This equation is, of course, entirely
equivalent to the master equation for the reduced density matrix and,
sometimes, we also refer to it as the master equation. The reduced density
matrix has been used to study decoherence induced by the environment \cite
{joos85,caldeira85,zurek93b,giulini96,unruh89,hu92,hu93}. The Wigner
function has also been used in studies of emergence of classicality induced
by an environment \cite{paz93b}, especially in quantum cosmology \cite
{habib90apaz91}.

The {\em Langevin equation} \cite{zwanzig73,sancho80} is another relevant
equation for open quantum systems. This equation has either been introduced
phenomenologically \cite{risken89} to describe the effect of the environment
into a classical system (Brownian motion) or it has been derived within the
functional approach (see, however, Refs. \cite{ford88} and \cite{gardiner91}
for a quantum version of the Langevin equation in operator language) as a
classical or semiclassical limit. Thus, Gell-Mann and Hartle \cite
{gell-mann93,hartle95} in the framework of the consistent histories approach
to quantum mechanics \cite{griffiths84} considered the decoherence
functional, which is closely related to the influence functional in the case
of open quantum systems, to measure the degree of classicality of the
system. They were able to show that under certain conditions there exists a
semiclassical limit in an open system which may be suitably described by a
Langevin equation with the self-correlation of the stochastic source given
by the noise kernel which appears in the decoherence functional.

Langevin-type equations as a suitable tool to study the semiclassical limit
have been used recently in semiclassical gravity and cosmology \cite
{calzetta94,martin99a,calzetta99}. In inflationary cosmology they have been
used to describe the stochastic effect on the inflaton field \cite
{starobinsky86,habib92a,calzetta95,calzetta97c,matacz97,polarski96,kiefer98}
or the stochastic behavior of large-scale gravitational perturbations \cite
{roura99b}, which is important for cosmological structure formation.

\subsection{A linear QBM model}

Here we review a QBM model as an example of a linear open quantum system.
Let us consider a harmonic oscillator of mass $M$, the ``system'', coupled
to a bath of independent harmonic oscillators of mass $m$, the
``environment''. For simplicity, let us assume that the system and
environment are linearly coupled. The action for the whole set of degrees of
freedom is defined by: 
\begin{equation}
S[x,\{q_{j}\}]=S[x]+S[\{q_{j}\}]+S_{int}[x,\{q_{j}\}]\text{,}  \label{2.1}
\end{equation}
where the terms on the right-hand side, which correspond to the action of
the system, the environment and the interaction term respectively, are given
by 
\begin{equation}
S[x]=\int dt(\frac{1}{2}M\dot{x}^{2}-\frac{1}{2}M\Omega ^{2}x^{2})\text{,}
\label{2.2}
\end{equation}
\begin{equation}
S[\{q_{j}\}]=\sum_{j}\int dt(\frac{1}{2}m\dot{q}_{j}^{2}-\frac{1}{2}m\omega
_{j}^{2}q_{j}^{2})\text{,}  \label{2.3}
\end{equation}
\begin{equation}
S_{int}[x,\{q_{j}\}]=\sum_{j}c_{j}\int dtx(t)q_{j}(t)=\int_{0}^{\infty
}d\omega \frac{2m\omega }{\pi c(\omega )}I(\omega ) \int dtx(t)q(t;\omega ) 
\text{,}  \label{2.4}
\end{equation}
where we introduced the spectral density $I(\omega )=\sum_{j}\pi
c_{j}^{2}(2m\omega _{j})^{-1}\delta (\omega -\omega _{j})$ in the last
equality, $c(\omega )$ and $q(t;\omega )$ are functions such that $c(\omega
_{j})=c_{j}$ and $q(t;\omega _{j})=q_{j}(t)$, with $c_{j}$ being
system-environment coupling parameters, and $\Omega $ and $\omega _{j}$ are,
respectively, the system and environment oscillator frequencies.

The reduced density matrix for an open quantum system is defined from the
density matrix $\rho $ of the whole system by tracing out the environment
degrees of freedom: 
\begin{equation}
\rho _r(x_f,x_f^{\prime },t_f)=\int \prod_jdq_j\rho (x_f,\{q_j\},x_f^{\prime
},\{q_j\},t_f)\text{.}  \label{2.5}
\end{equation}
The evolution for the reduced density matrix, which is in general nonunitary
and even non-Markovian, can be written as 
\begin{equation}
\rho _r(x_f,x_f^{\prime },t_f)=\int dx_idx_i^{\prime }J(x_f,x_f^{\prime
},t_f;x_i,x_i^{\prime },t_i)\rho _r(x_i,x_i^{\prime },t_i)\text{,}
\label{2.5b}
\end{equation}
where the propagator $J$ is defined in a path integral representation by 
\begin{equation}
J(x_f,x_f^{\prime },t_f;x_i,x_i^{\prime
},t_i)=\int\limits_{x(t_i)=x_i}^{x(t_f)=x_f}{\cal D}x\int\limits_{x^{\prime
}(t_i)=x_i^{\prime }}^{x^{\prime }(t_f)=x_f^{\prime }}{\cal D}x^{\prime
}e^{i(S[x]-S[x^{\prime }]+S_{IF}[x,x^{\prime }])/\hbar }\text{,}  \label{2.6}
\end{equation}
where $S_{IF}[x,x^{\prime }]$ is the influence action introduced by Feynman
and Vernon \cite{feynman63}. When the system and the environment are
initially uncorrelated, {\em i.e.}, when the initial density matrix
factorizes ($\hat \rho (t_i)=\hat \rho _r(t_i)\otimes \hat \rho _e(t_i)$,
where $\hat \rho _r(t_i)$ and $\hat \rho _e(t_i)$ mean, respectively, the
density matrix operators of the system and the environment at the initial
time) and the initial density matrix for the environment $\rho
_e(\{q_j^{(i)}\},\{q_j^{\prime (i)}\},t_i)$ is Gaussian, one obtains \cite
{feynman63,caldeira83a}: 
\begin{equation}
S_{IF}[x,x^{\prime }]=-2\int_{t_i}^{t_f}ds\int_{t_i}^sds^{\prime }\Delta
(s)D(s,s^{\prime })X(s^{\prime })+\frac i2\int_{t_i}^{t_f}ds
\int_{t_i}^{t_f}ds^{\prime }\Delta (s)N(s,s^{\prime }) \Delta (s^{\prime }) 
\text{,}  \label{2.8}
\end{equation}
where $X(s)\equiv (x(s)+x^{\prime }(s))/2$ and $\Delta (s)\equiv x^{\prime
}(s)-x(s)$. The kernels $D(s,s^{\prime })$ and $N(s,s^{\prime })$ are called
the dissipation and noise kernel, respectively. When the bath is initially
in thermal equilibrium these kernels are related by the usual
fluctuation-dissipation relation \cite{F-D}. When no special form is assumed
for the spectral density $I(\omega )$, this is usually referred to as a
general environment. One of the most common particular cases is the
so-called Ohmic environment, defined by $I(\omega )\sim \omega $ (some high
frequency cut-off may be sometimes naturally introduced).

Starting with Eqs.~(\ref{2.5}) and (\ref{2.6}) a differential equation for
the system's reduced density matrix, known as the master equation, can be
derived. The expression for a general environment was first obtained by Hu,
Paz and Zhang \cite{hu92}; see also Ref. \cite{halliwell96} for an
alternative derivation: 
\begin{eqnarray}
i\hbar \frac{\partial \rho _{r}}{\partial t} &=&-\frac{\hbar ^{2}}{2M}\left( 
\frac{\partial ^{2}}{\partial x^{2}}-\frac{\partial ^{2}}{\partial x^{\prime
2}}\right) \rho _{r}+\frac{1}{2}M\Omega ^{2}(x^{2}-x^{\prime 2})\rho _{r}+ 
\frac{1}{2}M\delta \Omega ^{2}(t)(x^{2}-x^{\prime 2})\rho _{r}  \nonumber \\
&&-i\hbar A(t)(x-x^{\prime })\left( \frac{\partial }{\partial x}- \frac{%
\partial }{\partial x^{\prime }}\right) \rho _{r}+ \hbar B(t)(x-x^{\prime
})\left( \frac{\partial }{\partial x}+ \frac{\partial }{\partial x^{\prime }}
\right) \rho _{r}-iMC(t)(x-x^{\prime })^{2}\rho _{r}\text{,}  \label{2.12}
\end{eqnarray}
where the functions $\delta \Omega ^{2}(t)$, $A(t)$, $B(t)$ and $C(t)$
represent a frequency shift, a dissipation factor and two diffusive factors,
respectively. For explicit expressions of these functions see Appendix \ref
{appB}. An alternative representation for the system reduced density matrix
is the reduced Wigner function $W_{r}(X,p,t)$ defined as 
\begin{equation}
W_{r}(X,p,t)=\frac{1}{2\pi \hbar }\int_{-\infty }^{\infty }d\Delta
e^{ip\Delta /\hbar }\rho _{r}(X-\Delta /2,X+\Delta /2,t)\text{.}
\label{2.13}
\end{equation}
It follows immediately that the master equation (\ref{2.12}) can be written
in the following equivalent form: 
\begin{equation}
\frac{\partial W_{r}}{\partial t}=\{H_{R},W_{r}\}_{PB}+2A(t)\frac{\partial
(pW_{r})}{\partial p}+\hbar B(t)\frac{\partial ^{2}W_{r}}{\partial X\partial
p}+\hbar MC(t)\frac{\partial ^{2}W_{r}}{\partial p^{2}}\text{,}  \label{2.14}
\end{equation}
where $\{H_{R},W_{r}\}_{PB}\equiv -(p/M)\partial W_{r}/\partial X+M\Omega
_{R}^{2}(t)X\partial W_{r}/\partial p$ with $\Omega _{R}^{2}(t)=\Omega
^{2}+\delta \Omega ^{2}(t)$. This equation is formally similar to the
Fokker-Planck equation for a distribution function.

\section{STOCHASTIC DESCRIPTION OF THE SYSTEM'S QUANTUM DYNAMICS}

\label{sec3}

In this Section we show that the reduced Wigner function can be written as a
formal distribution function for some stochastic process, and using this
result we deduce the corresponding Fokker-Planck equation.

\subsection{Reduced density matrix and Wigner function}

To find an explicit expression for the reduced density matrix (\ref{2.5}) at
a time $t_f$, we need to compute the path integrals which appear in Eq.~(\ref
{2.6}) for the reduced density matrix propagator. On the other hand, the
reduced Wigner function $W_r$ is related to the reduced density matrix by
the integral transform (\ref{2.13}). In Appendix \ref{appA}, we show that $%
W_r$ can be written in the following suggestive form: 
\begin{equation}
W_r(X_f,p_f,t_f)=\left\langle \left\langle \delta (X(t_f)-X_f)\delta (M\dot X
(t_f)-p_f)\right\rangle _\xi \right\rangle _{X_i,p_i}\text{,}  \label{wf}
\end{equation}
where 
\begin{eqnarray}
\left\langle \dots\right\rangle _\xi &\equiv &\left[ \det (2\pi N)\right]
^{- \frac 12}\int {\cal D}\xi \dots e^{-\frac 12\xi \cdot N^{-1}\cdot \xi } 
\text{,}  \label{Gauss} \\
\left\langle \dots \right\rangle _{X_i,p_i} &\equiv &\int_{-\infty }^\infty
dX_i\int_{-\infty }^\infty dp_i\dots W_r(X_i,p_i,t_i)\text{.}
\label{inaverage}
\end{eqnarray}
and $X\left( t\right) $ is the solution with initial conditions $X_i$, $p_i$
of the Langevin-like equation 
\begin{equation}
(L\cdot X)(t)=\xi (t)\text{,}  \label{3.14b}
\end{equation}
where $L(t,t^{\prime })\equiv M\left( \frac{d^2}{dt^{\prime 2}}+
\Omega_{ren}^2\right) \delta (t-t^{\prime })+H(t,t^{\prime })$. Here the
functions $\Omega_{ren}$ and $H(t,t^\prime)$ are defined below Eq.~(\ref{3.1}%
), and we have also used the notation $A\cdot B\equiv \int_{t_i}^{t_f}
dt\,A(t)B(t)$.

Thus, the reduced Wigner function can be interpreted as an average over a
Gaussian stochastic process $\xi (t)$ with $\left\langle \xi
(t)\right\rangle _\xi =0$ and $\left\langle \xi (t)\xi (t^{\prime
})\right\rangle _\xi =N(t,t^{\prime })$ as well as an average over the
initial conditions characterized by a distribution function $%
W_r(X_i,p_i,t_i) $. It is only after formally interpreting $\xi (t)$ as a
stochastic process characterized by Eq.~(\ref{Gauss}) that Eq.~(\ref{3.14b})
can be regarded as a Langevin equation. Note that, in general, Eq. (\ref
{3.14b}) is not meant to describe the actual trajectories of the system.

Note, in addition, that although $W_{r}(X_{i},p_{i},t_{i})$ is real, which
follows from the hermiticity of the density matrix, and properly normalized,
in general it is not positive everywhere (except for Gaussian states) and,
thus, cannot be considered as a probability distribution. The fact that the
Wigner function cannot be interpreted as a phase-space probability density
is crucial since most of the nonclassical features of the quantum state are
tightly related to the Wigner function having negative values. For instance,
a coherent superposition state is typically characterized by the Wigner
function presenting strong oscillations with negative values in the minima 
\cite{paz93b,giulini96}, which are closely connected to interference terms.

Equation (\ref{wf}) is the main result of this Section and shows that the
reduced Wigner function can be interpreted as a formal distribution in phase
space. This result will now be used to derive the corresponding
Fokker-Planck equation.

\subsection{From Langevin to Fokker-Planck: recovery of the master equation}

As mentioned above there is a simple one-to-one correspondence between any
density matrix and the associated Wigner function introduced in (\ref{2.13}%
). Taking this correspondence into account, the equation satisfied by the
reduced Wigner function is equivalent to the master equation satisfied by
the reduced density matrix. By deriving Eq.~(\ref{wf}) with respect to time
and using the Langevin-type equation in (\ref{3.14b}), one can obtain a
Fokker-Planck differential equation describing the time evolution of the
system's reduced Wigner function. The details of the calculation can be
found in Appendix \ref{appB}. Our result is the transport equation (\ref
{3.31}) which is written in terms of the time dependent coefficients $A(t)$, 
$B(t)$ and $C(t)$ defined, respectively, by Eqs.~(\ref{3.24}), (\ref{3.32})
and (\ref{3.33}). These coefficients are, of course, in agreement with those
previously derived in Refs. \cite{halliwell96} and \cite{hu92}; see also 
\cite{paz94}. Thus, this is yet another alternative way to derive the master
or transport equation (\ref{2.14}).

This new road to the transport equation highlights the fact that while one
can derive the Fokker-Planck equation from the Langevin equation, the
opposite is not possible in general. One can always consider Langevin
equations with stochastic sources characterized by different noise kernels
which, nevertheless, lead to the same Fokker-Planck equation and, thus, the
same master equation. This can be argued from the expressions obtained in
the derivation of the Fokker-Planck equation. Let us consider, for
simplicity, the situation corresponding to local dissipation. A local
contribution to the noise gives no contribution to $B(t)$, but it does
contribute to $C(t)$ as can be seen from Eqs.~(\ref{3.32}) and (\ref{3.33})
taking into account that $G_{ret}(t,t)=0$ and $\left. \partial
G_{ret}(t^{\prime },t)/\partial t^{\prime }\right| _{t^{\prime }=t}=M^{-1}$.
Thus, one can always choose any noise kernel that gives the desired $B(t)$
and then add the appropriate local contribution to the noise kernel to get
the desired $C(t)$ keeping $B(t)$ fixed. Note that changing the noise kernel
does not change $A(t)$. To illustrate the fact that there exist different
noise kernels giving the same $B(t)$, as was stated above, one may consider
the particular case corresponding to the weak dissipation limit so that $%
G_{ret}(t,t^{\prime })\sim (M\Omega )^{-1}\sin\Omega (t-t^{\prime })\theta
(t-t^{\prime })$. To see that a different $\tilde N(t,t^{\prime })$ giving
the same $B(t)$ as $N(t,t^{\prime })$ exists reduces then to show that there
is at least one nontrivial function $\nu (s,t)=$ $\tilde N(t,t^{\prime
})-N(t,t^{\prime })$ (with $s=t-t^{\prime }$) such that for any $t$ we have $%
\int_0^tds\sin (\Omega s)\nu (s,t)=0$, which can be shown to be the case.

The fact that different Langevin equations lead to the same master equation%
\footnote{%
In fact, what we showed was that a Langevin equation contains in general
more information that the corresponding Fokker-Planck equation. To extend
this assertion to the master equation, one should make sure that the
different Langevin equations leading to the same Fokker-Planck equation can
be obtained from an influence functional. This fact seems plausible provided
that one considers general Gaussian initial states for the environment.}
reflects that the former contains more information than the latter. This
fact can be qualitatively understood in the following way. In the influence
functional it is only the evolution of the environment degrees of freedom
that is traced out. Of course, having integrated over all the possible
quantum histories for the environment, no correlations in the environment
can be obtained. Nevertheless, since the system is interacting with the
environment, non-Markovian correlations for the system at different times
may in general persist. On the other hand, when considering either the
reduced density matrix or its propagator, also the system evolution, except
for the final state, is integrated out. Consequently, information on
non-Markovian correlations for the system is no longer available. Thus, only
when the system's reduced dynamics is Markovian, {\em i.e.} the influence
functional is local in time, we expect that the Langevin equation and the
master equation contain the same information. In particular, for a Gaussian
stochastic source, as in our case, the Langevin equation contains the
information about the system correlations at different times which the
Fokker-Planck equation cannot in general account for. Only in the case in
which the dynamics generated by the Langevin equation is Markovian one can
compute the correlation functions just from the solutions of the
Fokker-Planck equation or, equivalently, the master equation for the
propagator $J(x_{2},x_{2}^{\prime },t_{2};x_{1},x_{1}^{\prime },t_{1})$; see
Eq.~(\ref{2.6}). The key point is the fact that the propagator for the
reduced density matrix only factorizes when the influence action is local.
In Appendix \ref{appC} we give a detailed argument on this point.

It is important to note that for a closed quantum system the evolution
determined by the time evolution operators $U(t_{2},t_{1})$ obtained from
the Schr\"{o}dinger equation is always unitary and, thus, also Markovian.
That is why the Schr\"{o}dinger equation suffices to get the correlation
functions for a closed quantum system. On the contrary, for an open quantum
system the evolution is nonunitary and, provided that the influence action
is nonlocal, not even Markovian.

\section{CORRELATION FUNCTIONS}

\label{sec4}

We have seen that the reduced Wigner function, or equivalently the reduced
density matrix, and the master equation governing these functions can be
obtained from a formal stochastic description provided by the Langevin
equation (\ref{3.14b}). In this Section we show that also entirely quantum
correlation functions for the system can be obtained by means of the
stochastic description developed in the previous Section.

\subsection{CTP generating functional for the system and $n$-point quantum
correlation functions}

All the relevant quantum correlation functions for the system can be
obtained from the CTP generating functional, which is expressed, after
integrating out the environment, as \cite{ctp1,ctp2}: 
\begin{equation}
Z_{CTP}[J,J^{\prime }]=\int dx_f\int dx_idx_i^{\prime
}\int\limits_{x(t_i)=x_i}^{x(t_f)=x_f}{\cal D}x\int\limits_{x^{\prime
}(t_i)=x_i^{\prime }}^{x^{\prime }(t_f)=x_f}{\cal D}x^{\prime }e^{iJ\cdot
x-iJ^{\prime }\cdot x^{\prime }}e^{i(S[x]-S[x^{\prime }]+S_{IF}[x,x^{\prime
}])}\rho _r(x_i,x_i^{\prime },t_i)\text{,}  \label{3.39}
\end{equation}
where use was made of the influence action introduced in Eq.~(\ref{2.6}).
Equivalently, we may rewrite the previous equation changing to semisum and
difference variables with $J_\Sigma =(J(t)+J^{\prime }(t))/2$ and $J_\Delta
=J^{\prime }(t)-J(t)$, integrate the system action by parts and proceed
analogously as we did in Appendix A to obtain 
\begin{equation}
Z_{CTP}[J_\Sigma ,J_\Delta ]=\left\langle e^{-iJ_\Delta \cdot
X_o}\right\rangle _{X_i,p_i}e^{-\frac 12J_\Delta \cdot G_{ret}\cdot N\cdot
(J_\Delta \cdot G_{ret})^T}e^{-iJ_\Delta \cdot G_{ret}\cdot J_\Sigma } \text{%
.}  \label{3.43}
\end{equation}
where $X_0(t)$ is the solution to the homogeneous equation $(L\cdot X)(t)=0$
with the initial conditions $X_i$, $p_i$. It is interesting to note that the
first factor in Eq.~(\ref{3.43}) contains all the information about the
initial conditions of the system, whereas the information about the
fluctuations induced on the system by the environment is essentially
contained in the second factor through the noise kernel. This is the key
result of this Section, which will allow to relate the quantum with the
stochastic correlation functions.

Any $n$-point quantum correlation function for the system position operators
can be obtained from the CTP generating functional according to the equation 
\begin{equation}
Tr\left[ \left( T\hat x(t_1)\dots\hat x(t_m)\right) \hat \rho (t_i)\left( 
\tilde T\hat x(t_{m+1})\dots\hat x(t_{m+n})\right) \right] =i^{n-m}\left.
\left( \frac \delta {\delta J}\right) ^m\left( \frac \delta {\delta
J^{\prime }}\right) ^nZ_{CTP}[J,J^{\prime }]\right| _{J,J^{\prime }=0} \text{%
.}  \label{3.44a}
\end{equation}
Since one can always write $J$ and $J^{\prime }$ in terms of $J_\Sigma $ and 
$J_\Delta $, the right-hand side of Eq.~(\ref{3.44a}) can be expressed as a
linear combination of terms of the type 
\begin{equation}
i^{r+s}\left. \left( \frac \delta {\delta J_\Sigma }\right) ^r\left( \frac %
\delta {\delta J_\Delta }\right) ^sZ_{CTP}[J_\Sigma ,J_\Delta ]\right|
_{J_\Sigma ,J_\Delta =0}\text{,}  \label{3.44b}
\end{equation}
with $0\leq r\leq n+m$, $0\leq s\leq n+m$ and $r+s=n+m$. To obtain an
explicit expression one must evaluate Eq.~(\ref{3.44b}) with the final
result for the CTP generating functional (\ref{3.43}).

\subsection{Quantum correlation functions from stochastic averages}

Using the expression (\ref{3.44b}) for the case $r=0$, a connection can be
established between the correlation functions for the Gaussian stochastic
process associated to $\xi (t)$ via the Langevin-type equation (\ref{3.14b})
with $W_{r}(X_{i},p_{i},t_{i})$ as the distribution function for the initial
conditions, and some quantum correlation functions corresponding to quantum
expectation values of products of Heisenberg operators at different instants
of time. Any correlation function for the former stochastic process can be
obtained from its characteristic functional in the usual way: 
\begin{equation}
\left\langle \left\langle X(t_{1})\dots X(t_{s})\right\rangle _{\xi
}\right\rangle _{X_{i},p_{i}}=i^{s}\left. \left( \frac{\delta }{\delta K}%
\right) ^{s}\left\langle \left\langle e^{-i\,K\cdot X}\right\rangle _{\xi
}\right\rangle _{X_{i},p_{i}}\right| _{K=0}\text{.}  \label{3.62}
\end{equation}
The generating functional for the aforementioned stochastic process is, in
turn, related to the full CTP generating functional previously introduced as
follows: 
\begin{equation}
\left\langle \left\langle e^{-i\,K\cdot X}\right\rangle _{\xi }\right\rangle
_{X_{i},p_{i}}=Z_{CTP}[J_{\Sigma }=0,J_{\Delta }=K]\text{.}  \label{3.63}
\end{equation}
Substituting Eq.~(\ref{3.63}) into Eq.~(\ref{3.62}), rewriting $J_{\Delta }$
in terms of $J$ and $J^{\prime }$, and using expression (\ref{3.44a}), we
can express the correlation functions for the stochastic process in terms of
quantum correlation functions for the system observables. In particular, for 
$s=2$ we have 
\begin{equation}
\left\langle \left\langle X(t_{1})X(t_{2})\right\rangle _{\xi }\right\rangle
_{X_{i},p_{i}}=\frac{1}{4}\left[ \left\langle T\hat{x}(t_{1})\hat{x}%
(t_{2})\right\rangle +\left\langle \hat{x}(t_{1})\hat{x}(t_{2})\right\rangle
+\left\langle \hat{x}(t_{2})\hat{x}(t_{1})\right\rangle +\left\langle \tilde{%
T}\hat{x}(t_{1})\hat{x}(t_{2})\right\rangle \right] =\frac{1}{2}\left\langle
\left\{ \hat{x}(t_{1}),\hat{x}(t_{2})\right\} \right\rangle \text{,}
\label{3.48}
\end{equation}
where, as usual, we used $\left\langle \dots \right\rangle $ to denote the
quantum expectation value $Tr[\dots \hat{\rho}(t_{i})]$.

On the other hand, concentrating on the stochastic description provided by
the left-hand side of Eq.~(\ref{3.48}) and elaborating a little bit on it by
using Eq.~(\ref{3.6}) and taking into account that $\xi (t)$ is a Gaussian
stochastic process characterized by $\left\langle \xi (t)\right\rangle _{\xi
}=0$ and $\left\langle \xi (t_{1})\xi (t_{2})\right\rangle _{\xi
}=N(t_{1},t_{2})$, we can write 
\begin{eqnarray}
\left\langle \left\langle X(t_{1})X(t_{2})\right\rangle _{\xi }\right\rangle
_{X_{i},p_{i}} &=&\left\langle \left\langle \left[
X_{o}(t_{1})+(G_{ret}\cdot \xi )(t_{1})\right] \left[
X_{o}(t_{2})+(G_{ret}\cdot \xi )(t_{2})\right] \right\rangle _{\xi
}\right\rangle _{X_{i},p_{i}}  \nonumber \\
&=&\left\langle X_{o}(t_{1})X_{o}(t_{2})\right\rangle
_{X_{i},p_{i}}+(G_{ret}\cdot N\cdot (G_{ret})^T)(t_{1},t_{2})\text{.}
\label{3.49}
\end{eqnarray}
Hence, the final result is 
\begin{equation}
\frac{1}{2}\left\langle \left\{ \hat{x}(t_{1}),\hat{x}(t_{2})\right\}
\right\rangle =\left\langle X_{o}(t_{1})X_{o}(t_{2})\right\rangle
_{X_{i},p_{i}}+(G_{ret}\cdot N\cdot (G_{ret})^T)(t_{1},t_{2})\text{.}
\label{3.50}
\end{equation}
The left-hand side of Eq.~(\ref{3.50}) is the quantum correlation function,
which can therefore be described within the stochastic scheme in terms of
two separate contributions: the first term on the right-hand side
corresponds entirely to the {\em dispersion} in the initial conditions,
whereas the second term is due to the fluctuations induced by the stochastic
source appearing in the Langevin-type equation (\ref{3.14b}). It should be
remarked that, as discussed in Appendix \ref{appC}, for the general case of
a nonlocal influence action no quantum correlation functions (except for the
trivial case of $n=1$) can be expressed in terms of the propagators for the
reduced density matrix, which can be obtained from the master equation.

It is clear from Eqs.~(\ref{3.62}) and (\ref{3.63}) that only those quantum
correlation functions which are obtained by functionally differentiating the
CTP generating functional with respect to $J_{\Delta }$ (but not $J_{\Sigma
} $) an arbitrary number of times can be related to the stochastic
correlation functions (\ref{3.62}). Let us, therefore, see what is the
general expression for all the quantum correlation functions that can be
directly obtained from the stochastic description. We begin with the
classical correlation functions (\ref{3.62}) for the stochastic processes $%
X(t)$ which are solutions of the Langevin-type equation with stochastic
source $\xi (t)$ and initial conditions averaged over the initial reduced
Wigner function. Then we write these correlation functions in terms of path
integrals and use the results of Secs.~\ref{sec3} and \ref{sec4} to relate
them to a subclass of quantum correlation functions for the system: 
\begin{eqnarray}
\left\langle \left\langle X(t_{1})\dots X(t_{n})\right\rangle _{\xi
}\right\rangle _{X_{i},p_{i}} &=&\left[ \det (2\pi N)\right] ^{-\frac{1}{2}%
}\int_{-\infty }^{\infty }dX_{f}\int_{-\infty }^{\infty }dX_{i}\int_{-\infty
}^{\infty }dp_{i}\int {\cal D}\xi e^{-\frac{1}{2}\xi \cdot N^{-1}\cdot \xi }
\nonumber \\
&&\delta (X(t_{f})-X_{f})X(t_{1})\dots
X(t_{n})W_{r}(X_{i},p_{i}t_{i})=Tr^{*}\left[ \hat{X}(t_{1})\dots \hat{X}%
(t_{n})\hat{\rho}(t_{i})\right] \text{,}
\end{eqnarray}
with $\hat{X}(t_{j})=(\hat{x}(t_{j})+\hat{x}^{\prime }(t_{j}))/2$ and 
\begin{eqnarray}
Tr^{*}\left[ \hat{x}(t_{1})\dots \hat{x}^{\prime }(t_{r})\dots \hat{x}%
(t_{s})\dots \hat{x}^{\prime }(t_{u})\dots \hat{\rho}(t_{i})\right] &\equiv
&Tr\left[ \left\{ T\hat{x}(t_{1})\dots \hat{x}(t_{r-1})\hat{x}(t_{s})\dots 
\hat{x}(t_{u-1})\right\} \right.  \nonumber \\
&&\left. \hat{\rho}(t_{i})\left\{ \tilde{T}\hat{x}(t_{r})\dots \hat{x}%
(t_{s-1})\hat{x}(t_{u})\dots \right\} \right] \text{,}
\end{eqnarray}
where both the initial density matrix and the trace correspond to the whole
closed quantum system ({\em i.e.}, system plus environment) and $T$ and $%
\tilde{T}$ denote time and anti-time ordering, respectively. It is then
straightforward to show that 
\begin{equation}
\left\langle \left\langle X(t_{1})\dots X(t_{n})\right\rangle _{\xi
}\right\rangle _{X_{i},p_{i}}=2^{-n}\sum_{m=0}^{n}\frac{1}{m!(n-m)!}%
\sum_{\sigma \in S_{n}}Tr\left[ \left\{ T\prod_{j=\sigma (1)}^{\sigma (m)}%
\hat{x}(t_{j})\right\} \hat{\rho}(t_{i})\left\{ \tilde{T}\prod_{k=\sigma
(m+1)}^{\sigma (n)}\hat{x}(t_{k})\right\} \right] \text{,}
\end{equation}
where $\sigma \in S_{n}$ are all the possible permutations for a set
consisting of $n$ elements.

\section{SUMMARY AND DISCUSSION}

\label{sec5}

In this paper we have considered the stochastic description of a linear open
quantum system. We have shown that the reduced Wigner function can be
written as a formal distribution function for a stochastic process given by
a Langevin-type equation. The master equation has then been deduced as the
corresponding Fokker-Planck equation for the stochastic process. We have
also shown that a subclass of quantum correlation functions for the system
variables can be written in terms of stochastic correlation functions. Our
results are summarized in the diagram of Fig. 1 which show all the
interconnections between the influence functional, the Langevin equation,
the Fokker-Planck equation, the master equation and the correlation
functions.

Finally, we stress that although we have exploited the formal description of
open quantum systems in terms of stochastic processes, a classical
statistical interpretation is not always possible. Thus, although the Wigner
function is a real and properly normalized function providing a distribution
for the initial conditions of our formal stochastic processes, it is not a
true probability distribution function in the sense that it is not positive
definite in general. In fact, this property is crucial for the existence of
quantum coherence for the system. Nevertheless, even though the Langevin
equation does not in general describe actual classical trajectories
(histories) of the system, it is still a very useful tool to compute quantum
correlation functions or even as an intermediate step to derive the master
equation. Note that, after all, in statistical mechanics one uses Langevin
equations basically to compute correlation functions. It is remarkable that
in the light of our results the use of Langevin equations in semiclassical
gravity or in inflationary cosmology may provide more information about
genuine quantum properties of the gravitational field than previously
suspected, and its use as an intermediate step between the semiclassical
limit and a quantum theory seems justified.

\acknowledgements

We are grateful to Rosario Mart\'\i n for interesting discussions and to
Daniel Arteaga for a careful reading of the manuscript. This work has been
partially supported by the CICYT Research Project No. AEN98-0431 and by
Fundaci\'on Antorchas under grant A-13622/1-21. E.\ C.\ acknowledges support
from Universidad de Buenos Aires, CONICET, Fundaci\'on Antorchas and ANPCYT
through grant 03-05229. A.\ R.\ also acknowledges partial support of a grant
from the Generalitat de Catalunya, and E.\ V.\ also acknowledges support
from the Spanish Ministery of Education under the FPU grant PR2000-0181 and
the University of Maryland for hospitality.

\newpage

\appendix

\section{Derivation of the stochastic representation for the Wigner function}

\label{appA}

Let us begin by rewriting the influence action (\ref{2.8}) as 
\begin{eqnarray}
S_{IF}[x,x^{\prime }]
&=&\int_{t_{i}}^{t_{f}}ds\int_{t_{i}}^{t_{f}}ds^{\prime }\Delta
(s)H_{bare}(s,s^{\prime })X(s^{\prime })+\frac{i}{2}\int_{t_{i}}^{t_{f}}ds%
\int_{t_{i}}^{t_{f}}ds^{\prime }\Delta (s)N(s,s^{\prime })\Delta (s^{\prime
})  \nonumber \\
&\equiv &X\cdot H_{bare}\cdot \Delta +\frac{i}{2}\Delta \cdot N\cdot \Delta 
\text{,}  \label{3.1}
\end{eqnarray}
where we used the notation $A\cdot B\equiv \int_{t_{i}}^{t_{f}}dt\,A(t)B(t)$%
, and defined $H_{bare}(s,s^{\prime })$ as formally equivalent to $%
-2D(s,s^{\prime })\theta (s-s^{\prime })$. Being the product of two
distributions the latter expression is not well defined in general and
suitable regularization and renormalization may be required; see Ref. \cite
{roura99a} for details. The local divergences present in $%
H_{bare}(s,s^{\prime })=H(s,s^{\prime })+H_{div}(s)\delta (s-s^{\prime })$
can be canceled by suitable counterterms $\Omega _{div}$ in the bare
frequency of the system $\Omega =\Omega _{ren}+\Omega _{div}$. From now on
we will consider that this infinite renormalization, if necessary, has
already been performed so that both $\Omega _{ren}$ and $H(s,s^{\prime })$
are free of divergences. Now we perform three main steps.

First, we integrate the system action by parts: 
\begin{equation}
S[x]-S[x^{\prime }]=-M\int_{t_{i}}^{t_{f}}dt(\dot{X}(t)\dot{\Delta}%
(t)-\Omega _{ren}^{2}X(t)\Delta (t))=-\left. M\dot{X}\Delta \right|
_{t_{i}}^{t_{f}}+M\int_{t_{i}}^{t_{f}}dt\Delta (t)\left( \frac{d^{2}}{dt^{2}}%
+\Omega _{ren}^{2}\right) X(t)\text{.}  \label{3.2}
\end{equation}

Second, we perform the Gaussian path integral for $\Delta (t)$. Taking into
account that the value of the Jacobian determinant for the change of
integration variables $\int_{x_{i}}^{x_{f}}{\cal D}x\int_{x_{i}^{\prime
}}^{x_{f}^{\prime }}{\cal D}x^{\prime }\longrightarrow \int_{X_{i}}^{X_{f}}%
{\cal D}X\int_{\Delta _{i}}^{\Delta _{f}}{\cal D}\Delta $ is one, the
Gaussian path integral for $\Delta (t)$ with $\Delta _{i}$ and $\Delta _{f%
\text{ }}$ fixed is performed: 
\begin{equation}
\int_{X_{i}}^{X_{f}}{\cal D}X\int_{\Delta _{i}}^{\Delta _{f}}{\cal D}\Delta
e^{i\Delta \cdot L\cdot X}e^{-\frac{1}{2}\Delta \cdot N\cdot \Delta }=\left(
\det \frac{N}{2\pi }\right) ^{-\frac{1}{2}}\int_{X_{i}}^{X_{f}}{\cal D}Xe^{-%
\frac{1}{2}(L\cdot X)\cdot N^{-1}\cdot (L\cdot X)}\text{,}  \label{3.3}
\end{equation}
where $L(t,t^{\prime })\equiv M\left( \frac{d^{2}}{dt^{\prime 2}}+\Omega
_{ren}^{2}\right) \delta (t-t^{\prime })+H(t,t^{\prime })$. The integration
over $\Delta _{i}$ gives, 
\begin{eqnarray}
\rho _{r}(X_{f}-\Delta _{f}/2,X_{f}+\Delta _{f}/2,t_{f}) &=&\left( \det 
\frac{N}{2\pi }\right) ^{-\frac{1}{2}}\int_{-\infty }^{\infty }d\Delta
_{i}\int_{-\infty }^{\infty }dX_{i}\int_{X_{i}}^{X_{f}}{\cal D}Xe^{-\frac{1}{%
2}(L\cdot X)\cdot N^{-1}\cdot (L\cdot X)}  \nonumber \\
&&e^{-iM\dot{X}_{f}\Delta _{f}}e^{iM\dot{X}_{i}\Delta _{i}}\rho
_{r}(X_{i}-\Delta _{i}/2,X_{i}+\Delta _{i}/2,t_{i})  \nonumber \\
&=&2\pi \left( \det \frac{N}{2\pi }\right) ^{-\frac{1}{2}}\int_{-\infty
}^{\infty }dX_{i}\int_{X_{i}}^{X_{f}}{\cal D}Xe^{-\frac{1}{2}(L\cdot X)\cdot
N^{-1}\cdot (L\cdot X)}  \nonumber \\
&&e^{-iM\dot{X}_{f}\Delta _{f}}W_{r}(X_{i},M\dot{X}_{i},t_{i})\text{,}
\label{3.4}
\end{eqnarray}
where in the last step we used Eq.~(\ref{2.13}), which defines the reduced
Wigner function.

Third, we carry out the following functional change: 
\begin{equation}
X(t)\longrightarrow \left\{ X_{i}=X(t_{i})\text{, }p_{i}\equiv M\dot{X}_{i}=M%
\dot{X}(t_{i})\text{, }\xi (t)=(L\cdot X)(t)\text{ }\right\} \text{.}
\label{3.5}
\end{equation}
Note that with this change the function $X(t)$ gets substituted by the
initial conditions $(X_{i},p_{i})$ and the function $\xi (t)$ in the
functional integration. It is important to note that at this point the
function $\xi (t)$ is not a stochastic process but just a function over
which a path integral is performed. The functional change (\ref{3.5}) is
invertible as can be explicitly seen: 
\begin{equation}
\left\{ X_{i}\text{, }p_{i}\text{, }\xi (t)\right\} \longrightarrow
X(t)=X_{0}(t)+\int_{t_{i}}^{t}dt^{\prime }G_{ret}(t,t^{\prime })\xi
(t^{\prime })\text{,}  \label{3.6}
\end{equation}
where $G_{ret}(t^{\prime },t^{\prime \prime })$ is the retarded ({\em i.e.}, 
$G_{ret}(t^{\prime },t^{\prime \prime })=0$ for $t^{\prime }\leq t^{\prime
\prime }$) Green function for the linear integro-differential operator
associated to the kernel $L(t,t^{\prime })$, and $X_{i}(t)=$ $%
\int_{t_{i}}^{t}dt^{\prime }G_{ret}(t,t^{\prime })\xi (t^{\prime })$ is a
solution of the inhomogeneous equation $(L\cdot X_{i})(t)=\xi (t)$ with
initial conditions $X_{i}(t_{i})=0$ and $\left. \partial X_{i}(t^{\prime
})/\partial t^{\prime }\right| _{t^{\prime }=t_{i}}=0$. On the other hand, $%
X_{0}(t)$ is a solution of the homogeneous equation $(L\cdot X_{0})(t)=0$,
with initial conditions $X_{0}(t_{i})=X_{i}$ and $\dot{X}(t_{i})=p_{i}/M$.
Since the change is linear, the Jacobian functional determinant will be a
constant (this can be clearly seen by skeletonizing the path integral).
After performing the functional change, we obtain 
\[
\rho _{r}(X_{f}-\Delta _{f}/2,X_{f}+\Delta _{f}/2,t_{f})=K\int_{-\infty
}^{\infty }dX_{i}\int_{-\infty }^{\infty }dp_{i}\int {\cal D}\xi \delta
(X(t_{f})-X_{f})e^{-\frac{1}{2}\xi \cdot N^{-1}\cdot \xi }e^{-iM\dot{X}%
(t_{f})\Delta _{f}}W_{r}(X_{i},p_{i},t_{i})\text{,} 
\]
where the delta function $\delta (X(t_{f})-X_{f})$ was introduced to
restrict the functional integral $\int {\cal D}\xi $ with free ends, in
order to take into account the restriction on the final points of the
allowed paths for the integral $\int^{X_{f}}{\cal D}X$ appearing in Eq.~(\ref
{3.4}). The contribution from the Jacobian has been included in the constant 
$K$. In order to determine this constant, we demand the reduced density
matrix to remain normalized, {\em i.e.}, that $Tr\rho _{r}(t_{f})=1$ if $%
Tr\rho _{r}(t_{i})=1$: 
\begin{eqnarray}
1=\int_{-\infty }^{\infty }dX_{f}\rho _{r}(X_{f},X_{f},t_{f}) &=&K\int
dX_{f}\int {\cal D}\xi \delta (X(t_{f})-X_{f})e^{-\frac{1}{2}\xi \cdot
N^{-1}\cdot \xi }\int_{-\infty }^{\infty }dX_{i}\int_{-\infty }^{\infty
}dp_{i}W_{r}(X_{i},p_{i},t_{i})  \nonumber \\
&=&K\int {\cal D}\xi e^{-\frac{1}{2}\xi \cdot N^{-1}\cdot \xi }\int_{-\infty
}^{\infty }dX_{i}\int_{-\infty }^{\infty }dp_{i}W_{r}(X_{i},p_{i},t_{i})%
\text{.}  \nonumber
\end{eqnarray}
Now, from Eq.~(\ref{2.13}) it can be checked that $Tr\rho _{r}(t_{i})=1$
implies $\int_{-\infty }^{\infty }dX_{i}\int_{-\infty }^{\infty
}dp_{i}W_{r}(X_{i},p_{i},t_{i})=1$. The constant $K$ is thus determined to
be 
\begin{equation}
K=\left[ \int {\cal D}\xi e^{-\frac{1}{2}\xi \cdot N^{-1}\cdot \xi }\right]
^{-1}=\left[ \det (2\pi N)\right] ^{-\frac{1}{2}}\text{.}  \label{3.9}
\end{equation}

Finally, using the definition (\ref{2.13}) for the Wigner function and the
fact that 
\[
\frac{1}{2\pi }\int_{-\infty }^{\infty }d\Delta _{f}e^{ip_{f}\Delta
_{f}}e^{-iM\dot{X}(t_{f})\Delta _{f}}=\delta (M\dot{X}(t_{f})-p_{f})\text{,} 
\]
we get an expression for the reduced Wigner function 
\begin{equation}
W_{r}(X_{f},p_{f},t_{f})=K\int_{-\infty }^{\infty }dX_{i}\int_{-\infty
}^{\infty }dp_{i}\int {\cal D}\xi \delta (X(t_{f})-X_{f})\delta (M\dot{X}%
(t_{f})-p_{f})e^{-\frac{1}{2}\xi \cdot N^{-1}\cdot \xi
}W_{r}(X_{i},p_{i},t_{i})\text{,}  \label{3.11}
\end{equation}
which can be written as Eq.~(\ref{wf}).

\section{Derivation of the Fokker-Planck equation}

\label{appB}

The derivation of a Fokker-Planck equation from a Langevin equation with
local dissipation is well understood, see Ref. \cite{sancho80}. However, in
our case the existence of nonlocal dissipation makes it convenient to review
the main steps. Let us begin by computing $\partial W_{r}/\partial t$ from
expression (\ref{wf}), 
\begin{eqnarray}
\frac{\partial W_{r}(X,p,t)}{\partial t} &=&\left\langle \left\langle \dot{X}%
(t)\delta ^{\prime }(X(t)-X)\delta (M\dot{X}(t)-p)\right\rangle _{\xi
}\right\rangle _{X_{i},p_{i}}+\left\langle \left\langle \delta (X(t)-X)M%
\ddot{X}(t)\delta ^{\prime }(M\dot{X}(t)-p)\right\rangle _{\xi
}\right\rangle _{X_{i},p_{i}}  \nonumber \\
&=&-\frac{p}{M}\frac{\partial W_{r}(X,p,t)}{\partial X}-\frac{\partial }{%
\partial p}\left\langle \left\langle \delta (X(t)-X)M\ddot{X}(t)\delta (M%
\dot{X}(t)-p)\right\rangle _{\xi }\right\rangle _{X_{i},p_{i}}\text{,}
\label{3.15}
\end{eqnarray}
where the fact that $\dot{X}(t)$, $\partial /\partial X(t)$ and $\partial
/\partial \dot{X}(t)$ may be replaced by $p/M$, $-\partial /\partial X$ and $%
-\partial /\partial p$ respectively, since they are multiplying the delta
functions, was used in the second equality. Let us now concentrate on the
expectation value appearing in the last term and recall the expectation
values defined in (\ref{Gauss})-(\ref{inaverage}). We will consider the
Langevin-type equation 
\begin{equation}
(L\cdot X)(t^{\prime })=\xi (t^{\prime })\text{,}  \label{3.16}
\end{equation}
corresponding to the functional change (\ref{3.5}) and substitute the
corresponding expression for $M\ddot{X}(t)$ so that the last expectation
value in (\ref{3.15}) can be written as 
\begin{equation}
-M\Omega _{ren}^{2}XW_{r}(X,p,t)+\left\langle \left\langle \left(
-\int_{t_{i}}^{t}dtH(t,t^{\prime })X(t^{\prime })+\xi (t)\right) \delta
(X(t)-X)\delta (M\dot{X}(t)-p)\right\rangle _{\xi }\right\rangle
_{X_{i},p_{i}}\text{.}  \label{3.17}
\end{equation}
Any solution of Eq.~(\ref{3.16}) can be written as 
\begin{equation}
X(t^{\prime })=X_{h}(t^{\prime })+\int_{t_{i}}^{t}dt^{\prime \prime }\tilde{G%
}_{adv}(t^{\prime },t^{\prime \prime })\xi (t^{\prime \prime })\text{,}
\label{3.18}
\end{equation}
where $X_{h}(t^{\prime })$ is a solution of the homogeneous equation $%
(L\cdot X)(t^{\prime })=0$ such that $X_{h}(t)=X$, $\dot{X}_{h}(t)=p/M$ and $%
\tilde{G}_{adv}(t^{\prime },`t^{\prime \prime })$ is the advanced ({\em i.e.}
, $\tilde{G}_{adv}(t^{\prime },t^{\prime \prime })=0$ for $t^{\prime }\geq
t^{\prime \prime }$) Green function for the linear integro-differential
operator associated to the kernel $L(t,t^{\prime })$. The particular
solution of the inhomogeneous Eq.~(\ref{3.16}) $\tilde{X}_{i}(t^{\prime
})=\int_{t_{i}}^{t}dt^{\prime \prime }\tilde{G}_{adv}(t^{\prime },t^{\prime
\prime })\xi (t^{\prime \prime })$ has boundary conditions $\tilde{X}%
_{i}(t)=0$, $\left. \partial \tilde{X}_{i}(t^{\prime })/\partial t^{\prime
}\right| _{t^{\prime }=t}=0$. Both $X_{h}(t^{\prime })$ and $\tilde{G}%
_{adv}(t^{\prime },t^{\prime \prime })$ can be expressed in terms of the
homogeneous solutions $u_{1}(t^{\prime })$ and $u_{2}(t^{\prime })$, which
satisfy $u_{1}(t_{i})=1$, $u_{1}(t)=0$ and $u_{2}(t_{i})=0$, $u_{2}(t)=1$
respectively: 
\begin{eqnarray}
X_{h}(t^{\prime }) &=&X\left( u_{2}(t^{\prime })-\frac{\dot{u}_{2}(t)}{\dot{u%
}_{1}(t)}u_{1}(t^{\prime })\right) +\frac{(p/M)}{\dot{u}_{1}(t)}%
u_{1}(t^{\prime })\text{,}  \label{3.19} \\
\tilde{G}_{adv}(t^{\prime },t^{\prime \prime }) &=&-\frac{1}{M}\frac{%
u_{1}(t^{\prime })u_{2}(t^{\prime \prime })-u_{2}(t^{\prime
})u_{1}(t^{\prime \prime })}{\dot{u}_{1}(t^{\prime \prime })u_{2}(t^{\prime
\prime })-\dot{u}_{2}(t^{\prime \prime })u_{1}(t^{\prime \prime })}\theta
(t^{\prime \prime }-t^{\prime })\text{.}  \label{3.20}
\end{eqnarray}
We use the advanced propagator so that there is no dependence on the initial
conditions at time $t^{\prime }=t_{i}$ coming from the homogeneous solution
but just on the final conditions at time $t^{\prime }=t$, {\em i.e.}, on
those the Fokker-Planck equation is written in terms of. Using expression (%
\ref{3.18}) the first term within the expectation value appearing in Eq.~(%
\ref{3.17}) can be reexpressed as 
\begin{eqnarray}
&&\int_{t_{i}}^{t}dtH(t,t^{\prime })\left\langle \left\langle X(t^{\prime
})\delta (X(t)-X)\delta (M\dot{X}(t)-p)\right\rangle _{\xi }\right\rangle
_{X_{i},p_{i}}  \nonumber \\
&=&\int_{t_{i}}^{t}dt^{\prime }H(t,t^{\prime })X_{h}(t^{\prime
})W_{r}(X,p,t)+\int_{t_{i}}^{t}dt^{\prime }\int_{t_{i}}^{t}dt^{\prime \prime
}H(t,t^{\prime })\tilde{G}_{adv}(t^{\prime },t^{\prime \prime })\left\langle
\left\langle \xi (t^{\prime \prime })\delta (X(t)-X)\delta (M\dot{X}%
(t)-p)\right\rangle _{\xi }\right\rangle _{X_{i},p_{i}}\text{.}  \label{3.21}
\end{eqnarray}
The first term on the right-hand side can in turn be written as 
\begin{equation}
-\left( M\delta \Omega (t)X+2A(t)p\right) W_{r}(X,p,t)\text{,}  \label{3.22}
\end{equation}
where 
\begin{eqnarray}
\delta \Omega (t) &=&\frac{1}{M}\int_{t_{i}}^{t}dt^{\prime }H(t,t^{\prime
})[u_{2}(t^{\prime })-(\dot{u}_{2}(t)/\dot{u}_{1}(t))u_{1}(t^{\prime })]%
\text{,}  \label{3.23} \\
A(t) &=&\frac{1}{2}(M\dot{u}_{1}(t))^{-1}\int_{t_{i}}^{t}dt^{\prime
}H(t,t^{\prime })u_{1}(t^{\prime })\text{.}  \label{3.24}
\end{eqnarray}

In order to find an expression for $\left\langle \xi (t^{\prime })\delta
(X(t)-X)\delta (M\dot{X}(t)-p)\right\rangle _{\xi }$ we use Novikov's
formula for Gaussian stochastic processes \cite{novikov65}, which
corresponds essentially to use (\ref{Gauss}) and functionally integrate by
parts with respect to $\xi (t)$, 
\begin{equation}
\left\langle \xi (t^{\prime })F(t;\xi ]\right\rangle _{\xi
}=\int_{t_{i}}^{t}dt^{\prime \prime }N(t^{\prime },t^{\prime \prime
})\left\langle \delta F(t;\xi ]/\delta \xi (t^{\prime \prime })\right\rangle
_{\xi }\text{.}  \label{3.25}
\end{equation}
We then obtain the following expression: 
\begin{eqnarray}
\left\langle \xi (t^{\prime })\delta (X(t)-X)\delta (M\dot{X}%
(t)-p)\right\rangle _{\xi } &=&-\int_{t_{i}}^{t}dt^{\prime \prime
}N(t^{\prime },t^{\prime \prime })\left\langle \left( \frac{\delta X(t)}{%
\delta \xi (t^{\prime \prime })}\frac{\partial }{\partial X}+M\frac{\delta 
\dot{X}(t)}{\delta \xi (t^{\prime \prime })}\frac{\partial }{\partial p}%
\right) \right.  \nonumber \\
&&\left. \delta (X(t)-X)\delta (M\dot{X}(t)-p)\right\rangle _{\xi }
\label{3.26}
\end{eqnarray}
where we used again the presence of the delta functions to substitute the
functional derivatives $\delta /\delta X(t^{\prime \prime \prime })$ and $%
\delta /\delta \dot{X}(t^{\prime \prime \prime })$ by $\delta (t^{\prime
\prime \prime }-t)\cdot \partial /\partial X$ and $\delta (t^{\prime \prime
\prime }-t)\cdot M\cdot \partial /\partial p$, respectively, in the second
equality. Functionally differentiating with respect to $\xi (t^{\prime
\prime })$ expression (\ref{3.6}) for $X(t)$ and analogously for $\dot{X}(t)$
we get 
\begin{mathletters}
\begin{eqnarray}
\frac{\delta X(t^{\prime })}{\delta \xi (t^{\prime \prime })}
&=&G_{ret}(t^{\prime },t^{\prime \prime })\text{,}  \label{3.27} \\
\frac{\delta \dot{X}(t^{\prime })}{\delta \xi (t^{\prime \prime })} &=&\frac{%
\partial }{\partial t^{\prime }}G_{ret}(t^{\prime },t^{\prime \prime })\text{%
,}  \label{3.28}
\end{eqnarray}
which after substitution into (\ref{3.26}) leads to 
\end{mathletters}
\begin{equation}
\left\langle \left\langle \xi (t^{\prime })\delta (X(t)-X)\delta (M\dot{X}%
(t)-p)\right\rangle _{\xi } \right\rangle _{X_i,p_i}
=-\int_{t_{i}}^{t}dt^{\prime \prime }N(t^{\prime
},t^{\prime \prime })\left( G_{ret}(t,t^{\prime \prime })\frac{\partial }{%
\partial X}+M\frac{\partial G_{ret}(t,t^{\prime \prime })}{\partial t}\frac{%
\partial }{\partial p}\right) W_{r}(X,p,t)\text{.}  \label{3.29}
\end{equation}
The retarded Green function can also be expressed in terms of the solutions
of the homogeneous equation $u_{1}(t)$ and $u_{2}(t)$, which were previously
introduced, as 
\begin{equation}
G_{ret}(t^{\prime },t^{\prime \prime })=\frac{1}{M}\frac{u_{1}(t^{\prime
})u_{2}(t^{\prime \prime })-u_{2}(t^{\prime })u_{1}(t^{\prime \prime })}{%
\dot{u}_{1}(t^{\prime \prime })u_{2}(t^{\prime \prime })-\dot{u}%
_{2}(t^{\prime \prime })u_{1}(t^{\prime \prime })}\theta (t^{\prime
}-t^{\prime \prime })\text{.}  \label{3.30}
\end{equation}
Note that it is important to use now the expression in terms of the retarded
propagator $G_{ret}$ and the initial conditions $X_{i}$ and $p_{i}$ (at time 
$t^{\prime }=t_{i}$), since the ``final'' conditions $X(t)$ and $M\dot{X}(t)$
depend on $\xi (t^{\prime \prime })$ (for $t^{\prime \prime }<t$). Putting
all the terms together, {\em i.e.}, (\ref{3.17}), (\ref{3.21}) and (\ref
{3.29}), we reach the final expression for (\ref{3.15}): 
\begin{equation}
\frac{\partial W_{r}}{\partial t}=\{H_{R},W_{r}\}_{PB}+2A(t)\frac{\partial
(pW_{r})}{\partial p}+B(t)\frac{\partial ^{2}W_{r}}{\partial X\partial p}%
+MC(t)\frac{\partial ^{2}W_{r}}{\partial p^{2}}\text{,}  \label{3.31}
\end{equation}
where the Poisson bracket is defined following Eq.~(\ref{2.14}) (with $%
\Omega _{R}=\Omega _{ren}+\delta \Omega $), $\delta \Omega (t)$ and $A(t)$
are given by Eqs.~(\ref{3.23}) and (\ref{3.24}), and 
\begin{eqnarray}
B(t) &=&\int_{t_{i}}^{t}dt^{\prime \prime \prime }N(t,t^{\prime \prime
\prime })G_{ret}(t,t^{\prime \prime \prime })-\int_{t_{i}}^{t}dt^{\prime
}H(t,t^{\prime })\int_{t_{i}}^{t}dt^{\prime \prime }\tilde{G}%
_{adv}(t^{\prime },t^{\prime \prime })\int_{t_{i}}^{t}dt^{\prime \prime
\prime }N(t^{\prime \prime },t^{\prime \prime \prime })G_{ret}(t,t^{\prime
\prime \prime })\text{,}  \label{3.32} \\
C(t) &=&\int_{t_{i}}^{t}dt^{\prime \prime \prime }N(t,t^{\prime \prime
\prime })\frac{\partial G_{ret}(t,t^{\prime \prime \prime })}{\partial t}%
-\int_{t_{i}}^{t}dt^{\prime }H(t,t^{\prime })\int_{t_{i}}^{t}dt^{\prime
\prime }\tilde{G}_{adv}(t^{\prime },t^{\prime \prime
})\int_{t_{i}}^{t}dt^{\prime \prime \prime }N(t^{\prime \prime },t^{\prime
\prime \prime })\frac{\partial G_{ret}(t,t^{\prime \prime \prime })}{%
\partial t}\text{.}  \label{3.33}
\end{eqnarray}
The last two expressions were obtained by combining the second term within
the expectation value appearing in (\ref{3.17}) and the second term on the
right-hand side of Eq.~(\ref{3.21}). It should be taken into account that if
we put back the $\hbar $'s, there appears one with every noise kernel in
Eqs.~(\ref{3.32}) and (\ref{3.33}).

\section{Correlation functions and nonlocal influence action}

\label{appC}

Let us see how the fact that the influence action is nonlocal implies that
the propagator for the reduced density matrix does not factorize in time
and, thus, the system evolution is non-Markovian. In this Appendix we will
denote the integrand of the real part of the influence action by ${\cal H}%
\equiv \Delta (t)H(t,t^{\prime })X(t^{\prime })$ and the integrand of the
imaginary part by ${\cal N}\equiv \Delta (t)N(t,t^{\prime }) \Delta
(t^{\prime })/2$.

When the influence action is local ${\cal H}(t,t^{\prime })\equiv {\rm 
\tilde{H}}(t)\delta (t-t^{\prime })$, ${\cal N}(t,t^{\prime })\equiv {\rm 
\tilde{N}}(t)\delta (t-t^{\prime })$ and we have 
\begin{equation}
S_{IF}[x,x^{\prime
};t_{f},t_{i})=\int_{t_{i}}^{t_{f}}dt\int_{t_{i}}^{t_{f}}dt^{\prime }{\cal H}%
+i\int_{t_{i}}^{t_{f}}dt\int_{t_{i}}^{t_{f}}dt^{\prime }{\cal N}%
=\int_{t_{i}}^{t_{f}}dt{\rm \tilde{H}}+i\int_{t_{i}}^{t_{f}}dt{\rm \tilde{N}}%
\text{,}  \label{C1}
\end{equation}
where we introduced the notation $S_{IF}[x,x^{\prime };t_{f},t_{i})$, which
is a functional of $x(t)$ and $x^{\prime }(t)$ and also depends on the
variables $t_{i}$ and $t_{f}$, to explicitly state the initial and final
times defining the dependence domain considered for the functions $x(t)$ and 
$x^{\prime }(t)$, which will play an important role in the subsequent
discussion. Expression (\ref{C1}) can then be decomposed as follows 
\begin{equation}
S_{IF}[x,x^{\prime };t_{f},t_{i})=\left( \int_{t_{1}}^{t_{f}}dt{\rm \tilde{H}%
}+i\int_{t_{1}}^{t_{f}}dt{\rm \tilde{N}}\right) +\left(
\int_{t_{i}}^{t_{1}}dt{\rm \tilde{H}}+i\int_{t_{i}}^{t_{1}}dt{\rm \tilde{N}}%
\right) =S_{IF}[x,x^{\prime };t_{f},t_{1})+S_{IF}[x,x^{\prime };t_{1},t_{i})%
\text{,}  \label{C2}
\end{equation}
so that the influence functional factorizes 
\begin{equation}
F_{IF}[x,x^{\prime };t_{f},t_{i})=e^{iS_{IF}[x,x^{\prime
};t_{f},t_{i})}=F_{IF}[x,x^{\prime };t_{f},t_{1})F_{IF}[x,x^{\prime
};t_{1},t_{i})\text{,}  \label{C3}
\end{equation}
and so does the reduced density matrix propagator, as can be
straightforwardly seen from its path integral representation 
\begin{eqnarray}
J(x_{f},x_{f}^{\prime },t_{f};x_{i},x_{i}^{\prime },t_{i})
&=&\int\limits_{x(t_{i})=x_{i}}^{x(t_{f})=x_{f}}{\cal D}x\int\limits_{x^{%
\prime }(t_{i})=x_{i}^{\prime }}^{x^{\prime }(t_{f})=x_{f}^{\prime }}{\cal D}%
x^{\prime }e^{i\left( S[x]-S[x^{\prime }]+S_{IF}[x,x^{\prime
};t_{f},t_{i})\right) }  \nonumber \\
&=&\int dx_{1}dx_{1}^{\prime }\left(
\int\limits_{x(t_{1})=x_{1}}^{x(t_{f})=x_{f}}{\cal D}x\int\limits_{x^{\prime
}(t_{1})=x_{1}^{\prime }}^{x^{\prime }(t_{f})=x_{f}^{\prime }}{\cal D}%
x^{\prime }e^{i\left( S[x]-S[x^{\prime }]+S_{IF}[x,x^{\prime
};t_{f},t_{1})\right) }\right)  \nonumber \\
&&\times \left( \int\limits_{x(t_{i})=x_{i}}^{x(t_{1})=x_{1}}{\cal D}%
x\int\limits_{x^{\prime }(t_{i})=x_{i}^{\prime }}^{x^{\prime
}(t_{1})=x_{1}^{\prime }}{\cal D}x^{\prime }e^{i\left( S[x]-S[x^{\prime
}]+S_{IF}[x,x^{\prime };t_{1},t_{i})\right) }\right)  \nonumber \\
&=&\int dx_{1}dx_{1}^{\prime }J(x_{f},x_{f}^{\prime
},t_{f};x_{1},x_{1}^{\prime },t_{1})J(x_{1},x_{1}^{\prime
},t_{1};x_{i},x_{i}^{\prime },t_{i})\text{,}  \label{C4}
\end{eqnarray}
where use was made both of the fact that the system action is local and (\ref
{C3}) applied to definition (\ref{2.6}) for the reduced density matrix
propagator. This property allows one to obtain the quantum correlation
functions for the system from the propagators of the reduced density matrix,
which are solutions of the master equation. To illustrate this fact,
consider as an example the quantum correlation function $\left\langle \hat{x}%
(t_{2})\hat{x}(t_{1})\right\rangle $ with $t_{2}>t_{1}$, defined by: 
\begin{eqnarray}
Tr\left[ \hat{x}(t_{2})\hat{x}(t_{1})\hat{\rho}(t_{i})\right]
&=&\int\limits^{x(t_{f})=x_{f}}{\cal D}x\int\limits^{x^{\prime
}(t_{f})=x_{f}^{\prime }}{\cal D}x^{\prime }x(t_{2})x(t_{1})e^{i\left(
S[x]-S[x^{\prime }]+S_{IF}[x,x^{\prime };t_{f},t_{i})\right) }\rho
_{r}(x_{i},x_{i}^{\prime },t_{i})  \nonumber \\
&=&\int dx_{i}dx_{i}^{\prime }\int dx_{2}dx_{2}^{\prime }x_{2}\int
dx_{1}dx_{1}^{\prime }x_{1}J(x_{f},x_{f}^{\prime },t_{f};x_{2},x_{2}^{\prime
},t_{2})  \nonumber \\
&&J(x_{2},x_{2}^{\prime },t_{2};x_{1},x_{1}^{\prime
},t_{1})J(x_{1},x_{1}^{\prime },t_{1};x_{i},x_{i}^{\prime },t_{i})\rho
_{r}(x_{i},x_{i}^{\prime },t_{i})\text{.}  \label{C5}
\end{eqnarray}
Here the path integrals in the intermediate steps were decomposed in a way
completely analogous to that used in (\ref{C4}). Hence, the information on
the correlation functions can be essentially obtained from the master
equation when the influence action is local.

On the other hand, when the influence action is nonlocal, 
\begin{eqnarray}
S_{IF}[x,x^{\prime };t_{f},t_{i})
&=&\int_{t_{i}}^{t_{f}}dt\int_{t_{i}}^{t_{f}} dt^{\prime }{\cal H}
+i\int_{t_{i}}^{t_{f}}dt\int_{t_{i}}^{t_{f}}dt^{\prime }{\cal N}  \nonumber
\\
&=&\left( \int_{t_{i}}^{t_{1}}dt\int_{t_{i}}^{t_{1}} dt^{\prime }{\cal H}
+\int_{t_{i}}^{t_{1}}dt\int_{t_{1}}^{t_{f}} dt^{\prime }{\cal H}
+\int_{t_{1}}^{t_{f}}dt\int_{t_{i}}^{t_{1}} dt^{\prime }{\cal H}
+\int_{t_{1}}^{t_{f}}dt\int_{t_{1}}^{t_{f}} dt^{\prime }{\cal H}\right) 
\nonumber \\
&&+i\left( \int_{t_{i}}^{t_{1}}dt\int_{t_{i}}^{t_{1}} dt^{\prime }{\cal N}
+\int_{t_{i}}^{t_{1}}dt\int_{t_{1}}^{t_{f}} dt^{\prime }{\cal N}
+\int_{t_{1}}^{t_{f}}dt\int_{t_{i}}^{t_{1}} dt^{\prime }{\cal N}
+\int_{t_{1}}^{t_{f}}dt\int_{t_{1}}^{t_{f}} dt^{\prime }{\cal N}\right) 
\text{.}  \label{C6}
\end{eqnarray}
The cross terms like $\int_{t_{i}}^{t_{1}}dt\int_{t_{1}}^{t_{f}} dt^{\prime }%
{\cal N}$ do not allow the influence action to be separated into terms that
depend either on the ``history'' of the system just for times smaller than $%
t_{1}$ or just for times greater than $t_{1}$ (as happened in Eq.~(\ref{C2}%
)). This fact makes it impossible to factorize the influence functional as
was done in Eq.~(\ref{C3}) and consequently implies that neither the reduced
density matrix propagators factorize in the sense of Eq.~(\ref{C4}) nor the
quantum correlation functions can be obtained from the reduced density
matrix propagators as was done in Eq.~(\ref{C5}). It is, thus, clear how the
nonlocality of the influence action leads to a non-Markovian evolution for
the system and the impossibility to obtain the correlation functions from
the propagators for the reduced density matrix.

\end{document}